\documentclass[psfig]{ws-procs10x7}
\usepackage{graphics}
\usepackage{epsfig}
\usepackage{balance}

\newcolumntype{d}[1]{D{.}{.}{#1}}

\def\Journal#1#2#3#4{{\it #1} {\bf #2}, #3 (#4)}

\makeindex
\begin{document}

\title{EXPERIMENTAL RESULTS ON SEARCHES\\
BEYOND THE STANDARD MODEL}

\author{ELISABETTA GALLO}

\address{INFN Firenze, Via G. Sansone 1, Sesto Fiorentino (FI), 50019 Italy\\$^*$E-mail: gallo@mail.desy.de}


\twocolumn[\maketitle\abstract{Recent results for direct searches
for physics beyond the Standard Model are reviewed.
The results include Tevatron II data up to $1.2~\mathrm{fb}^{-1}$ and
HERA results up to $350~\mathrm{pb}^{-1}$. Searches for Supersymmetry,
for compositeness and for large
extra dimensions are presented. The excess of events with
an isolated lepton and high missing transverse momentum at HERA
is discussed.}
]

\section{Introduction}

We know from the corrections to the Higgs mass that new
physics could be at a scale of 1 or few TeV. It would be possible
then to see effects beyond the Standard Model (SM) at present colliders,
and of course in future at the LHC. 
In this report, recent results from direct
searches at HERA (up to $350~\mathrm{pb}^{-1}$ of integrated luminosity) 
and Tevatron (up to  $1.2~\mathrm{fb}^{-1}$) are reviewed.
A review
of the present theoretical ideas and phenomenological models 
beyond the SM can be found in~\cite{theorytalk}.

Searches for supersymmetry (SUSY) scenarios are reviewed in Sec. 2; searches
for compositeness models (like leptoquarks and
finite quark radius at HERA) are described in Sec. 3.
Models based on large extra-dimensions and Randall-Sundrum gravitons
are discussed in Sec. 4, which also describes searches for new heavy
gauge bosons.
Finally Sec. 5 describes a special signature studied at HERA, with isolated
electrons or muons accompanied by large missing transverse momentum.
Many more results can be found in the parallel session contributions
at this conference. Searches for Higgs and minimal supersymmetric Standard
Model (MSSM) Higgs at Tevatron and LHC
are reported in other plenary contributions~\cite{glenzinskifabiola}. 

\section{Search for Supersymmetry}

Supersymmetry is one of the most attractive model for new physics,
as it provides a solution to the hierarchy problem, provides unification
of the three interactions at the GUT scale  and a possible
candidate for the dark matter (the light supersymmetric particle, LSP).
For every particle, it predicts a partner (sparticle) with spin
differing by one half. As we know that the selectron mass is different
from the electron mass, Supersymmetry is a broken symmetry and
various scenarios have been proposed. One popular scenario
is the gravity mediated SUSY breaking scenario mSUGRA, where the $>100$
parameters in SUSY are reduced to only five parameters that can
completely describe the phenomenology: the common scalar mass at the GUT scale
($m_0$), the common gaugino mass ($m_{1/2}$), the common trilinear
soft trilinear SUSY breaking parameter ($A_0$), the ratio of
the Higgs vacuum expectation values to the electroweak scale ($\tan\beta$)
and the sign of the Higgsino mass term (sign $\mu=\pm 1$). 

Particles are assigned an 
$R$-parity $R_P=(-1)^{(3B+L+2S)}$=+1, while sparticle have $R_P$=-1,
where $R_P$ is a multiplicative new quantum number. In models where
$R_P$ is conserved, sparticles are produced in pairs and at the end
of the chain decay in the LSP (in many models
the lightest neutralino,$\tilde{\chi}_1^0$), which is weakly
interacting  and
escapes detection. This leads to a distinct signature of large missing
transverse momentum, for instance at the Tevatron. Always at the Tevatron
the cross sections are small, i.e. only $< \simeq$ 100 hundred events are expected for
$1~\mathrm{fb}^{-1}$ for various sparticles with masses under 400 GeV, but not impossible.
The Standard Model background must however be modelled and understood 
very well, this will be especially true at the LHC.

\subsection{Search for chargino and neutralino at the Tevatron}\label{subsec:chneutr}

\begin{figure}
\centerline{\epsfig{file=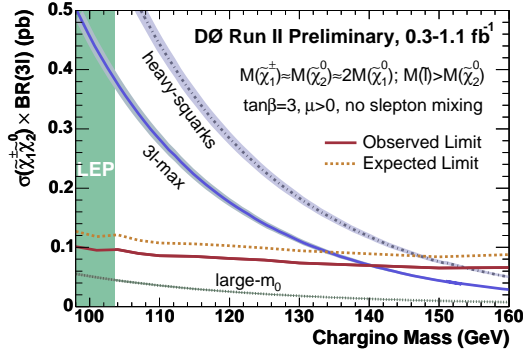,width=2.7in}}
\caption{Limits at $95\%$ CL on the $\sigma \times BR(3{l})$ for associated production
of the lightest neutralino and the second lightest chargino in the trilepton channel. 
The limit is derived from the combination of different channels in the final state
with integrated luminosities between 0.3 and 1.1 $\mathrm{fb}^{-1}$.
The limit of 140 GeV on the $\tilde{\chi}_1^\pm$ mass assumes 
the favourable mSUGRA $3l$-max scenario, where the sleptons are
mass degenerate, with mass just above the $\tilde{\chi}_2^0$ mass. }
\label{d03leptons}
\end{figure}

The charginos ($\tilde{\chi}_{1,2}^\pm$), 
which are the superpartner of the states coming from
the mixing of the $W^{\pm}$ and $H^{\pm}$, and the 
neutralinos ($\tilde{\chi}_{1,2,3,4}^0$),
which are the superpartners that come from the mixing of the neutral Higgses
and SU(2)$\times$U(1) bosons, are best searched at Tevatron in associated production.
The golden signature comes from the MSSM  channel $q \bar q \rightarrow W^* \rightarrow
\tilde{\chi}_1^+ \tilde{\chi}_2^0 \rightarrow {l} {l} {l} \nu \tilde{\chi}_1^0 \tilde{\chi}_1^0$. The final state has therefore three leptons and missing transverse energy (MET) coming from the neutrino and lightest neutralino, which is assumed to be the LSP.
The cross section is small but the process is basically
background free. Both CDF and D0 found no evidence in this signature, setting
limits on the lightest chargino, as shown in Fig.~\ref{d03leptons} for the D0 experiment.  A mass up to 140 GeV~\cite{d0cdfchneutr} on the $\tilde{\chi}_{1}^\pm$ mass is excluded, extending the
previous limit by LEP which was 103 GeV.

\begin{figure}
\centerline{\epsfig{file=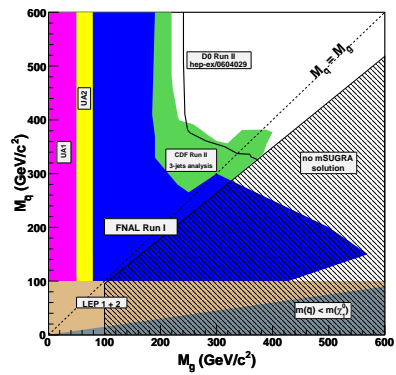,width=2.2in}}
\caption{Limits at $95\%$ CL on the quark and gluino mass from D0 (310$~\mathrm{pb}^{-1}$)
and CDF (371$~\mathrm{pb}^{-1}$). Note that the limits were derived with slighlty
different assumptions on $\tan\beta$ (=3 for D0, =5 for CDF). }
\label{fig1}
\end{figure}

\subsection{Search for RPC squarks at Tevatron}\label{subsec:squarks}

The inclusive production of squarks and gluinos is
one of the most promising discovery channels for SUSY at the Tevatron, 
due to the large cross section.
In $R_P$ conserving (RPC) scenarios, the cascade decays of the produced squark and gluino 
give a final state with quark, gluons and the LSP (the $\tilde{\chi}_1^0$), with
a distinct topology with multijets and missing $E_T$.
D0 and CDF have searched~\cite{d0cdfsqgl} for this signal optimizing the analyses
for the three cases in which the mass of the squark is greater, equal or smaller
than the mass of the gluino.
As no particular excess is observed, limits are set in the
two-dimensional plane   $M(\tilde{q}), M(\tilde{g})$ in the context of the
mSUGRA model. The result is shown in
Fig.~\ref{fig1}, where it can be seen that limits are set for  
$M(\tilde{q})>325~\mathrm{GeV}$ and $M(\tilde{g})>241~\mathrm{GeV}$ (D0) or
$M(\tilde{q,g})>387~\mathrm{GeV}$ (CDF) in the case of equal masses.
At LHC, squarks
and gluinos can be discovered with a significance of 5-$\sigma$ with $1~\mathrm{fb}^{-1}$ 
up to a mass of 1.5 TeV.

\begin{figure}
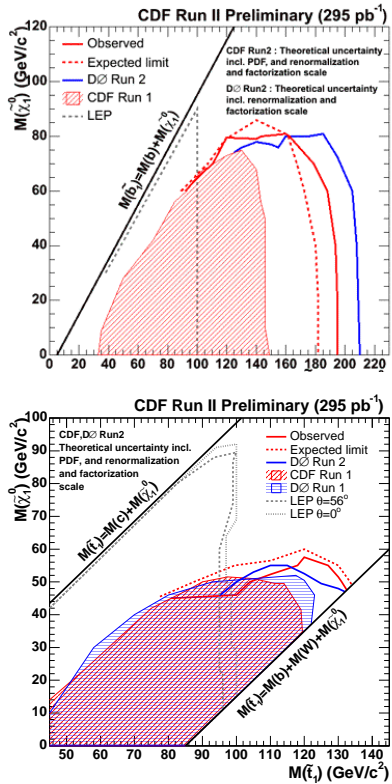

\centerline{\epsfig{file=gallo-plenary_3a.epsi,width=2in}}
\centerline{\epsfig{file=gallo-plenary_3b.epsi,width=2in}}
\caption{Exclusion limits at $95\%$ CL for the sbottom (upper plot) and stop mass (lower plot) as a function of the neutralino mass from D0 (310$~\mathrm{pb}^{-1}$)
and CDF (295$~\mathrm{pb}^{-1}$). }
\label{stoptevatron}
\end{figure}

In the case of the third generation squarks, there could be a large
mixing between the right-handed and left-handed weak eigenstates, leading
to a light mass eigenstate ($\tilde{t}_1,\tilde{b}_1$). 
In $R_P$-conserving models, sbottom 
and stop are produced in pairs via $q \bar q$ annihilation or gluon-gluon fusion
and the cross-section has  little dependence on the 
SUSY parameters other than the sbottom and stop mass,
respectively. 
The produced sbottom or stop decay then in a quark and the LSP,
$p \bar{p} \rightarrow  \tilde{t}_1 \bar{\tilde{t}}_1 \rightarrow
c  \tilde{\chi}_1^0 \bar{c} \tilde{\chi}_1^0$ and
$p \bar{p} \rightarrow  \tilde{b}_1 \bar{\tilde{b}}_1 \rightarrow
b  \tilde{\chi}_1^0 \bar{b} \tilde{\chi}_1^0$. The topology is 
two acoplanar charm- or b-jets with high missing $E_T$, and the main
background is W/Z+jets production. In the case of the stop search, 8 events were 
observed by D0 at $MET>150$~GeV with an expectation of 3 events from 
the SM. However the missing $E_T$ distribution of these events is larger than
expected for a stop signal. A detailed investigation
of these events did not reveal any anomaly. 

In both searches~\cite{d0cdfstsb}, then, no clear anomaly above the background
expectations was observed.
Limits on the sbottom mass up to 200 GeV were set as shown in Fig.~\ref{stoptevatron}, 
depending on the
neutralino mass, extending considerably the Run I limit, which
was approximately 150 GeV, and the limits from LEP. The exclusion area for the stop
is also shown in Fig.~\ref{stoptevatron} for the
stop mass versus the neutralino mass. The stop is excluded up
to 131 GeV, in an exclusion area complementary to LEP results.

\subsection{Search for events with photons and GMSB signatures}\label{subsec:gmsb}

\begin{figure}
\centerline{\epsfig{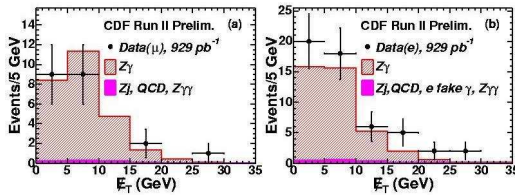}}
\caption{
Missing $E_T$ distribution for events with two muons or two electrons and
a photon in the CDF Run II data
(929$~\mathrm{pb}^{-1}$). }
\label{fig3}
\end{figure}

In Run I, CDF observed~\cite{cdfevent} 
a spectacular event with two $e$, two $\gamma$
and a missing $E_T$ of 55 GeV, to be compared to a SM expectation of
$10^{-6}$ events. A possible explanation of this event 
was $pp \rightarrow \tilde{e} \tilde{e} \rightarrow e \tilde{\chi}_2^0
e  \tilde{\chi}_2^0$, with $ \tilde{\chi}_2^0 \rightarrow
\gamma  \tilde{\chi}_1^0$. 
In addition 16 events with a lepton+photon+missing $E_T$ were observed for
an expectation of $7.6 \pm 0.7$. This has motivated a search~\cite{pronko} for events
with leptons, photons and missing energy also in Run II and the
analysis was repeated with the data taken
recently, based on $929~\mathrm{pb}^{-1}$.  In total 163 events, with an
expectation of $148.1 \pm 13.0$ were observed with the topology lepton+photon+$MET$. No
excess was also observed in events with either two muons or two electrons and 
a photon: the missing $E_T$ distribution for these events, shown in 
figure~\ref{fig3}, is always $<30~\mathrm{GeV}$ and no event with two photons
was observed. 

\begin{figure}
\centerline{\epsfig{file=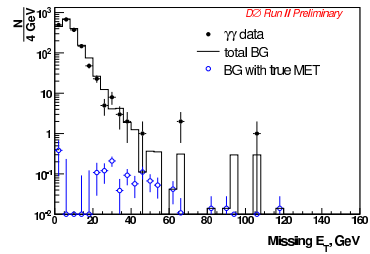,width=2in}}
\caption{
Missing $E_T$ distribution for events with two photons in $760~\mathrm{pb}^{-1}$ of 
D0 Run II data. The histogram represents the total background, 
which also includes QCD background, with either real photons or jets misidentified as
real photons, without true missing $E_T$.
The open circles represent
the background from events with true missing $E_T$, like $W (\rightarrow e \nu) \gamma$ and $W (\rightarrow e \nu) jet$, where the jet or the electron are misidentified as the second
photon. }
\label{diphoton}
\end{figure}

The observation of the CDF Run I event has also motivated 
D0 to perform a general search~\cite{d0gmsb} for events with multiphoton and missing $E_T$, a
topology which is expected in gauge mediated SUSY breaking scenarios (GMSB), 
where the gravitino is the LSP, the
neutralino is the next LSP, and decays in a gravitino and a photon. 
The missing transverse energy spectrum for events 
with two central high transverse energy photons 
($|\eta_\gamma|<1.1$ and $E_{T \gamma}>25$ GeV), is shown in Fig.~\ref{diphoton}. For missing
transverse energy greater than 45 GeV, there are 4 events observed, with an expectation
of $2.1\pm0.7$ events expected from QCD processes and $W \gamma,W+$jet processes. The process
is then used to set a limit on the GMSB scale $\Lambda$, which also gives the scale
of the gaugino masses: $\Lambda>88.5$~TeV, $M(\tilde\chi_1^0)>120$~GeV and 
$M(\tilde\chi_1^+)>220$~GeV.

\subsection{Search for RPV stop at HERA}\label{subsec:stop}

Sparticles can only be produced singly at HERA, via $R$-parity violating (RPV) reactions. 
As the stop is not yet well constrained at the Tevatron, it can be produced at HERA 
in the $s$-channel from the positron beam and a $d$-valence quark in the
proton. The cross-section is proportional to the square 
of the $\lambda^\prime_{131}$ $R_P$-violating coupling.
The stop can be detected as a narrow resonance, either decaying to $e^+d$ (electron-jet), 
like a leptoquark; or via the gauge decay $\tilde{t} \rightarrow b \tilde{\chi}_1^+$ 
and subsequent decays, giving topologies with multijet and
neutrino, or multijet and positron in the final state. 
These three channels
were recently investigated by ZEUS in the HERA I data~\cite{bellagamba}, 
observing no deviation
from the SM expectation of deep inelastic scattering (DIS) events. 
The limit set on the stop
mass, for a coupling similar to the electromagnetic strength
 of $\lambda^\prime_{131}=0.3$, 
is around 260 GeV. In the region of MSSM space where $M_2$ and $\mu$ are small, and therefore
the gauge decays dominate over the leptoquark-type decay,
the HERA limits are the most stringent to date.

\subsection{Search for stopped gluino in split-SUSY models}

Recently a new variant of SUSY, called split-Supersymmetry, has been proposed,
in which scalars are heavy, possibly at the GUT scale, compared to the SUSY
fermions~\cite{splitsusy}. Due to the high mass of the scalar particles, gluino
decays are suppressed and the gluinos have time to hadronize into `$R$-hadrons',
bound states of the gluino and  quarks or gluons. These coulourless $R$-hadrons live
long enough to reach the calorimeter and, in case they are charged, they can lose
all their momentum via ionization and come at rest in the calorimeters (so called
`stopped gluinos'). These stopped gluinos then decay in a gluon 
(giving a jet) and a neutralino (the LSP, giving missing transverse energy). 

\begin{figure}
\centerline{\epsfig{file=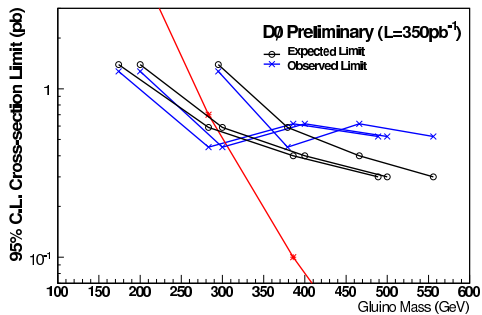,width=2.7in}}
\caption{The $95\%~C.L.$ limit on the cross-section for stopped
gluino decaying in a jet plus neutralino. The different curves correspond to the expected
and observed limit for 3 different values of the neutralino mass (from left to
right: 50, 90 and 200 GeV). The continous curve is the theoretical cross section.}
\label{sgluino}
\end{figure}

D0 has
performed a search~\cite{d0sgluino} for these $R$-hadrons: 
assuming that the stopped gluino has a
long enough lifetime (at least 10 $\mu s$), its decay would occur in
a bunch-crossing later than the one when it was produced. The signature is then a 
largely empty event with a high-$p_T$ jet and large missing $E_T$. The background is 
mainly cosmic muons and beam-halo muons. The observed spectrum of monojets for empty
events is in good agreement with the expected background. The D0 analysis excluded
gluino masses up to $\simeq 270~\mathrm{GeV}$, for a light neutralino, 
as shown in Fig.~\ref{sgluino}.

\section{Search for Compositeness}

\subsection{Search for leptoquarks}\label{subsec:res}

Leptoquarks (LQs) are colour triplet boson which appear naturally in
various unifying theories beyond the SM. At HERA, LQs
can be singly produced from the fusion of the
initial state electron/positron beam with a quark from the incoming
proton. For masses below the centre of mass energy of 318 GeV they
are produced as narrow resonances in the $s$-channel, for higher masses
they are exchanged in the $u$-channel and contribute to the DIS cross-section.

\begin{figure}
\centerline{\epsfig{file=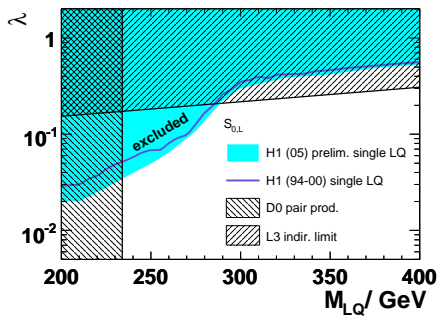,width=2.5in}}
\caption{
Exclusion Limits at $95\%~C.L.$ for the leptoquark $S_{0,L}$ in the BRW model. The
shaded area is obtained with the H1 data from HERA II. The indirect limits
from L3 and the direct limit from D0 are also shown.}
\label{lqlimits}
\end{figure}

H1 has presented a new analysis~\cite{h1lq} on searches for $F=2$ leptoquarks - 
where the fermion number $F$ is defined from the baryonic and leptonic numbers 
as $F=|3B+L|$-  profiting from the
recent large sample of $e^-p$ collisions of the 2005 data, where there is most
sensitivity to this type of leptoquarks. The signature is a narrow resonance
in the invariant mass of the electron and jet or neutrino and jet final-state 
system. The lepton
decay distribution in the center-of-mass frame of the hard subprocess can be
used to distinguish the signal from the DIS charged current (CC) and neutral current 
(NC) background. As no evidence above
the DIS background was found, H1 set limits on the LQ mass as a function of
the Yukawa coupling $\lambda$ of the LQ to the electron and quark. 
The limit is shown for instance for the LQ $S_{0,L}$ (leptoquarks with
F=2  have been classified in 7 different types by B\"uchmuller-R\"uckl-Wyler according to spin, isospin, chirality, hypercharge) in Fig.~\ref{lqlimits}.
H1 excludes, for a coupling of electromagnetic strength, leptoquark masses
in the range of 276-304 GeV. The plot shows also the nice complementarity
in this search between the colliders. At high masses the most stringent 
constraint comes from the indirect limits of L3 from the $e^+ e^- \rightarrow q \bar q$ cross-section. At low masses, the LQ mass is constrained below
234 GeV at all values of the coupling by the pair production in the D0~\cite{d0lq}
Run I+II data.

\subsection{Search for finite quark radius at HERA}\label{subsec:ci}

\begin{figure}
\centerline{\epsfig{file=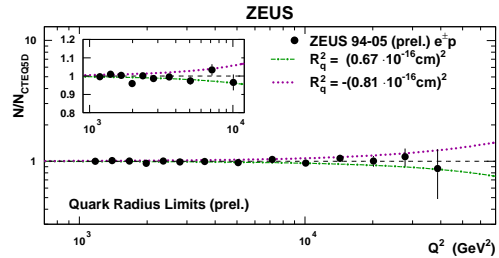,width=2.7in}}
\caption{
Ratio of the measured to predicted cross section as a function of
$Q^2$ for the 94-05 ZEUS data ($274~\mathrm{pb^{-1}}$). 
The SM prediction is calculated using the CTEQ5 proton 
parton-densities parametrization. The dashed-dotted line
correspond to the $95\%~C.L.$ limit for an effective mean-square radius of
the electroweak charge of the quark of $0.67 \cdot 10^{-16}~\mathrm{cm}$.  }
\label{zeusradius}
\end{figure}

The high $Q^2$ region, where $Q^2$ is the virtuality of the exchanged
boson, is the most interesting kinematic region to look for new physics at HERA,
especially for new interactions between electrons and quarks. If the scale of
these new interactions is above the center of mass energy ($\sqrt{s}\simeq 318~\mathrm{GeV}$),
the SM cross section can be modified via virtual effects. 
A finite charge radius for the quark would modify the $d\sigma/dQ^2$ with a form
factor type term of the form $(1- R^2_q/6 Q^2)^2$, where $R_q$ is the root-mean-square radius
of the electroweak charge of the quark. Figure~\ref{zeusradius}
shows the new preliminary results from ZEUS, using the combined HERAI+HERAII
dataset~\cite{zeusci}. As good agreement is shown between the data distribution
and the SM prediction, limits on the quark radius of 
$R_q < 0.67 \cdot 10^{-16}~\mathrm{cm}$ are set. This constraint is the most stringent 
to date.   

\section{Search for LED, $Z^\prime$ and $W^\prime$}

\subsection{Search for Large Extra Dimensions and Randall-Sundrum gravitons}

Models based on Large Extra Dimensions (LED), such as the model of Arkani-Hamed,
Dimopolous and Dvali (ADD)~\cite{addmodel}, offer a suitable framework
to solve the hierarchy problem. In these models  gravity is allowed to propagate in the
4+$n$ dimensional bulk of space-time, while the remainder of the SM fields are confined 
to the 3+1 dimensional world-volume of the brane configuration. The extra $n$ dimensions are
compactified with a radius $R$ and
the scale $M_D$ of gravity is related to the Planck scale by the relation
$M^2_{Pl}\simeq R^n M_D^{2+n}$. If the size of the extra-dimensions is of the order
of $\simeq 10 \mu m$, $M_D$ could be as small as the size of the weak scale (of the order of 1 TeV) and its effect could be visible at present colliders.

At HERA, after summing the effects of gravitons excitations in the extra-dimensions, the
graviton contribution could be visible as a contact interaction contributing to
the $eq \rightarrow eq$ scattering.
The effect could therefore be visible in the DIS neutral current cross-section, introducing
terms of the form $\eta_G= \lambda/M_S^4$ to the SM lagrangian, where $M_S$ is an ultraviolet cutoff scale, of the
order of $M_D$, and $\lambda=\pm 1$. The ZEUS experiment has presented new results based on
the same sample of data as used in Fig.~\ref{zeusradius}. As the cross-section shows good agreement
with the SM prediction versus $Q^2$, limits~\cite{zeusci} on the scale $M_S>0.88~\mathrm{TeV}, \lambda=+1$ and
$M_S>0.86~\mathrm{TeV}, \lambda=-1$ were derived.

\begin{figure}
\centerline{\epsfig{file=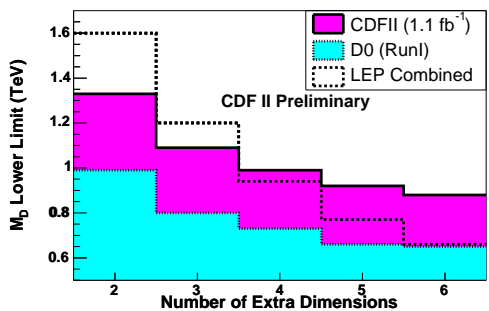,width=2.7in}}
\caption{
Limits on the scale of the compactified large extra dimensions
as a function of the number of extra dimensions. The CDF
results from Run II data ($1.1~\mathrm{fb}^{-1}$) are compared to limits
from D0 in Run I and from LEP. The LEP limits are extracted in a
search for events with single photon and missing energy.}
\label{ledlimit}
\end{figure}

At Tevatron (or LHC), gravitons can be produced directly in processes like
$q \bar q \rightarrow gG, q g \rightarrow q G, gg \rightarrow gG$ 
and escape undetected in the bulk of extra-dimensions, 
leaving a signature of a single energetic jet and large missing transverse energy. 
CDF~\cite{cdfled} has looked for this monojet signature 
in $1.1~\mathrm{fb}^{-1}$ of data. Spectacular events
like monojets with a transverse energy of $384~\mathrm{GeV}$ 
and missing $E_T$ of $390~\mathrm{GeV}$
are found. However the number of candidate events 
is in agreement with the expectation of the SM
background, which is expected mainly from $Z+jets$ production, 
with the $Z$ decaying in $\nu \bar{\nu}$. CDF set therefore
limits on the scale $M_D$ as a function of the number of extra-dimensions $n$, which
are
shown in Fig.~\ref{ledlimit}. For $n=2,3$, the most stringent limit on the scale (up 
to $1.6~\mathrm{TeV}$) is set by the LEP combined result, derived from searches for events with single photon and missing energy~\cite{lepsingle}. For $n \geq 4$,
CDF sets the most stringent limit, up to $M_D \simeq 1$ TeV. 

\begin{figure}
\centerline{\epsfig{file=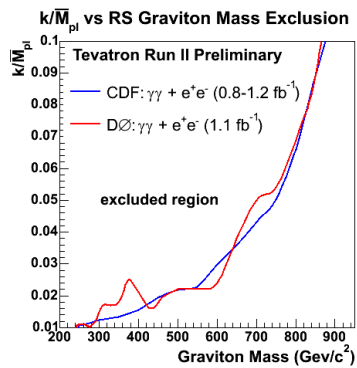,width=2in}}
\caption{
Limits at $95\%~C.L.$ on the mass of the first excited state of the graviton
in the Randall-Sundrum model, as a function of the dimensionless
coupling $k/\bar{M_{Pl}}$ of the model.
The results from CDF correspond to a combined limit  
in a search in $\gamma \gamma$ final states
with $\simeq 1.1~\mathrm{fb}^{-1}$ and 
in $e^+e^-$ final states with $\simeq 0.8~\mathrm{fb}^{-1}$. The D0
results are from a combined limit from $\gamma \gamma$ and $e^+e^-$ final states with
 $1.1~\mathrm{fb}^{-1}$.}
\label{rslimit}
\end{figure}

In the Randall-Sundrum (RS) model~\cite{randsund} gravity propagates in one
extra-dimension, which is highly curved (`warped'). The RS relationship between
the weak and the Planck scales is given by $M_W= e^{-2 k \pi R} M_{Pl}$.
In the model gravitons appear as Kaluza-Klein (KK) tower of massive excitations
and can be resonantly produced in $p \bar p$ collisions. The zeroth KK mode
remains massless and couples with the SM fields with gravitational strength,
while the excited modes couple with a strength which depends on the
curvature parameter $k$ and the scale $\bar{M}_{Pl}=M_{Pl}/\sqrt{8 \pi}$ ($k/\bar{M}_{Pl}$, which is expected
to be between 0.01 and 0.1) and decay as narrow resonances in fermion-antifermion
or diboson pairs. Both CDF and  D0 have performed~\cite{cdfd0rs,cdfzprime}
a search for the first excited mode of the KK graviton, looking for narrow
resonances with spin=2 in the dielectron and diphoton spectra.
Both the diphoton and the dilepton spectra are in agreement with the SM expectations.
The derived limits on the graviton mass
as a function of the coupling $k/\bar{M}_{Pl}$ are shown in Fig.~\ref{rslimit}. 
For $k/\bar{M}_{Pl}=0.1$
limits on the first excited mode of the graviton up to 875 (865) GeV for CDF (D0) are set. 
For $k/\bar{M}_{Pl}=0.01$, the limit on the RS graviton mass is around 230 GeV. The CMS Collaboration
has studied the discovery  potential for a graviton decaying in the 
channels $\gamma \gamma,ee, \mu \mu$: for a coupling of 0.1, masses up to 
3 TeV could be reached with $10~\mathrm{fb}^{-1}$ of integrated luminosity.

\subsection{Search for $Z^\prime$ and $W^\prime$ at Tevatron}

\begin{figure}
\centerline{\epsfig{file=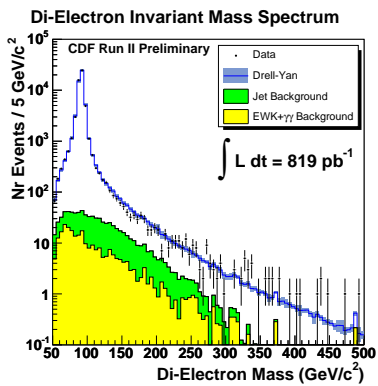,width=2in}}
\caption{
Invariant mass of electron pairs in the CDF Run II data, compared to the SM
expectation. }
\label{cdfdielectron}
\end{figure}

Many extensions of the Standard Model predict new heavy objects decaying to
electron pairs, like new heavy bosons $Z^\prime$.
The generic search for resonances decaying into $e^+e^-$ 
just described in the previous section can also then be applied 
to searches~\cite{cdfzprime} for spin-1 objects, like the $Z^\prime$. 
The dielectron mass spectrum in the CDF data is
shown in the range from 50 to 500 GeV in Fig.~\ref{cdfdielectron}. 
The search for new narrow resonances  is done up to
$>\simeq 900~\mathrm{GeV}$, but there are no events observed outside the scale
of the plot, with the highest mass candidate having a mass of 491 GeV. 
The spectrum is
perfectly reproduced by the SM contributions, mainly Drell-Yan production. Assuming that 
the $Z^\prime$ couples to the electrons in the same way as the Standard Model electroweak 
$Z$ boson, CDF set
a limit on the $Z^\prime$ mass of 850 GeV.  At the LHC, 
a new heavy boson $Z^\prime$ with a mass of 1 TeV could be discovered 
in the decays in $ee,\mu \mu$ with the first $0.1~\mathrm{fb}^{-1}$ accumulated.

\begin{figure}
\centerline{\epsfig{file=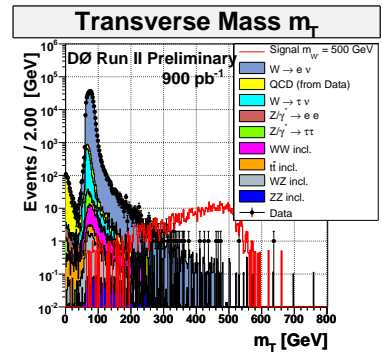,width=2in}}
\caption{
Transverse  mass in the electron channel  in the D0 Run II data, compared to the SM
expectation. The possible signal from a $W^\prime$ of mass of 500 GeV is also shown. }
\label{d0transversemass}
\end{figure}

The D0 Collaboration~\cite{d0wprime} has searched instead for new heavy charged gauge bosons, $W^\prime$. The new $W^\prime$ is supposed to have the same couplings to the fermions as that of the SM $W$ and to have a width scaling
with its mass.  The new decay channel $W^\prime \rightarrow WZ$ is assumed to be 
suppressed and new heavy leptons are assumed to be too heavy to be produced by a $W^\prime$. 
With these assumptions, the search
is performed in the $W^\prime \rightarrow e \nu$ channel. Events with a central electron with transverse energy of at least 30 GeV 
and large missing transverse energy are selected. 
The transverse mass spectrum of the sample is shown in Fig.~\ref{d0transversemass} in the range up to 800 GeV. The spectra is well described by the various SM backgrounds,
the main contribution being events from the
Standard Model W production, $W \rightarrow e \nu X$.
For a transverse mass above 150 GeV, 630 events are observed for an expectation of 623. D0 set therefore limits on the $W^\prime$ production, excluding its mass up to 965 GeV.

\section{Isolated lepton events at HERA}

\begin{figure}
\centerline{\epsfig{file=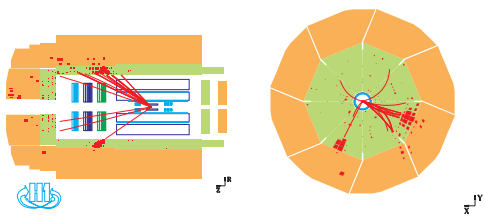,width=2.5in}}
\caption{
Event display of one of the isolated lepton events in H1. }
\label{h1event}
\end{figure}

\begin{figure}
\centerline{\epsfig{file=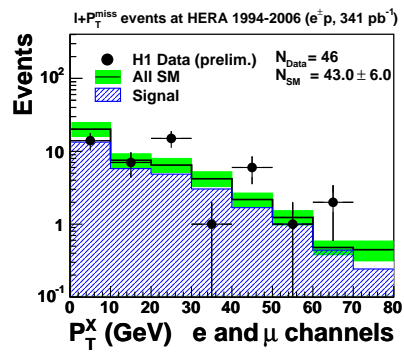,width=2.2in}}
\caption{
The $P_T^X$ distribution for the events with isolated leptons
in H1, corresponding to 341~$\mathrm{pb}^{-1}$ of $e^\pm p$ data. }
\label{h1ptx}
\end{figure}

Since 1994, the H1 Collaboration at HERA has observed particular
events, characterized by a high $p_T$ lepton (electron/positron or muon),
an hadronic system $X$ with high transverse energy $P_T^X$ and
high missing transverse momentum. One of these events
is shown in Fig.~\ref{h1event}. These events are expected
in the SM from $W$-production, where the $W$ is radiated from
the incoming quark and decays in an electron or muon and neutrino,
giving the topology mentioned above. The $P_T^X$ distribution for the
H1 1994-2006 data is shown in Fig.~\ref{h1ptx},  which shows in general
agreement between the data and the SM prediction, which
is dominated by $W$-production. However it can also be seen that at high $P_T^X$
there is a clear excess of observed events.

The number of events at high $P_T^X$~\cite{h1isol},
 $P_T^X>25~\mathrm{GeV}$, observed and expected by H1, is reported in 
Table~\ref{tab1}, separated in the $e^+p$ and in the $e^-p$ datasets. It is clearly
observed that the excess is in the positron-proton collisions:
in total H1 observes in $e^+p$ 15 events with an expectation of $4.6 \pm 0.8$, which is
a $3.4~\sigma$ excess. The effect is reduced to $2.6~\sigma$ taking into account also the
ZEUS events~\cite{zeusisol}, where a good agreement with the SM is observed in both channels, $e$ or $\mu$,
and in both datasets (see table). If the rate of these events would remain constant in H1,
a $4.5~\sigma$ effect could be reached with additional $100~\mathrm{pb}^{-1}$ of
$e^+p$ data. HERA has just restarted to collide positron and protons and plans
to continue until March 2007 in order  to collect the required amount of data.
The last months of HERA data-taking, until June 2007, will instead be devoted to a low
energy run to measure the longitudinal structure function. 

\begin{table}
\tbl{Number of events with isolated leptons in H1 (94-06) and ZEUS (98-05) data. All
numbers are preliminary. \label{tab1}}
{\begin{tabular}{@{}ccc@{}}
\toprule
{\small  $P_T^X >$ 25 GeV } & {\small  $e$} & {\small  $\mu$ } \\
\colrule
{\small  H1  $e^{-}p$ (184$\mathrm{pb}^{-1}$)  } & {\small 3/3.8} & {\small 0./3.1} \\ 
{\small  H1  $e^{+}p$ (158$\rm{pb}^{-1}$) } & {\small 9/2.3} & {\small 6./2.3} \\
\colrule
{\small  ZEUS $e^{-}p$ (143$\mathrm{pb}^{-1}$) } & {\small 3/2.9} & {\small 2./1.6} \\
{\small  ZEUS $e^{+}p$ (106$\mathrm{pb}^{-1}$) } &  {\small 1/1.5} & {\small 1./1.5} \\
\botrule
\end{tabular}}
\end{table}

\begin{figure}
\centerline{\epsfig{file=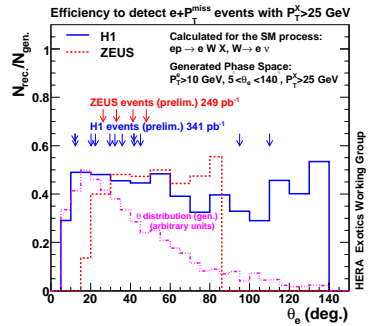,width=2in}}
\caption{
Efficiency for selecting $W \rightarrow e \nu$  events in H1 (solid line) and in
ZEUS (dashed line),
as a function of the polar angle of the electron in the final state.
The arrows indicate the electron polar angle for the events selected
at high $P_T^X$. The dotted curve shows the generated distribution.}
\label{h1zeuseff}
\end{figure}

In view of the discrepancy between the two experiments, recently~\cite{diaconu} the H1
and ZEUS Collaborations have compared their efficiencies for a benchmark process
for this type of topology, the 
$W \rightarrow e \nu, \mu \nu$ processes. An example is shown in
Fig.~\ref{h1zeuseff} where the acceptances for ZEUS and H1 are shown as
a function of the polar angle of the final state electron (for the muon channel,
it is very similar). The main differences in the two analyses is actually
in the range of this polar angle: as it can be seen it is restricted in ZEUS,
compared to H1. However, in the
region where they overlap, the two efficiencies are very similar. In addition most
of the interesting events in H1, shown by the arrows in the plot, lie in the
region of common acceptance. The ZEUS experiment plans to increase the efficiency
also in the forward region, where most of the decay products of the $W$ lie,
as shown by the dotted curve, and where new physics is expected.

\section{Summary}

No convincing sign of new physics was reported at this conference.
Tevatron did not confirm special events found in Run I: 
they have
 just started the second part of their Run II program (Run IIb) and
approximately $4\div8~\mathrm{pb}^{-1}$ are expected until 2009. HERA has just started
the last part of the HERA II run and expects to collect $>100~\mathrm{pb}^{-1}$ of $e^+p$
collisions until the end of HERA data-taking, which will be end of June 2007. 
This amount of integrated luminosity should be sufficient to give a definite
answer on the isolated leptons events observed by H1. Finally we are all waiting
eagerly for the start of LHC.

\section*{Acknowledgments}
I would like to thank the organizers for the invitation to give this talk
and the Tevatron, HERA, LHC and LEP  colleagues who helped me in the preparation.
Special thanks go to Emmanuelle Perez and Cristinel Diaconu (H1), Beate Heinemann (CDF) and 
Jean-Francois Grivaz (D0) for providing me the information from their
experiments and for their precious comments. 

\balance

\end{document}